\documentclass[epj]{svjour}
\usepackage{latexsym}
\usepackage{amssymb}
\usepackage{amsmath}
\usepackage{amscd}
\usepackage{graphicx}
\usepackage{textcomp}
\usepackage{colortbl}
\usepackage{subcaption}
\usepackage{hyperref}
\usepackage{multirow}
\usepackage{lipsum}
\begin{document}
	
\title{Note on Tsallis holographic dark energy in Brans-Dicke cosmology}
\author{Anil Kumar Yadav}
\institute{Department of Physics, United College of Engineering and Research,Greater Noida - 201310, India\\\textcolor{blue}{abanilyadav@yahoo.co.in}}


\abstract{
In this comment, we have investigated that the power law variation of Brans-Dicke scalar field $(\phi)$ with scale factor $(a)$ is not validated by field equations in general for the Universe filled with dark matter and Tsallis type holographic dark energy. However, for specific choice of free parameters $(\omega = -\frac{3}{2}\;\; \&\;\; n = -1)$ which are differ from its values fixed in Ghaffari et al \textcolor{blue}{[Ghaffari et al. Euro. Phys. J. C \textbf{78}, 706 (2018)]} to describe various physical properties of the Universe in derived model, the power law variation of $\phi$ with $a$ leads to consistent results. Further, we notice that the energy conservation law for Tsallis holographic dark energy in Brans-Dicke gravity does not hold. 
}

\authorrunning{A. K. Yadav}
\titlerunning{Note on Tsallis holographic dark energy}

\maketitle

\section{Introduction}
\label{sec:intro}
The Brans - Dicke theory of gravity \cite{Brans/1961,Clifton/2012} is the simplest extension of Einstein's General Theory of Relativity depending on one additional parameter $\omega$ - dimensionless coupling parameter between scalar field and gravity. In addition to the space-time manifold, the gravitational field is further mediated by a scalar field $\phi(t)$ which plays the role of inverse of space-time varying gravitational strength $G(t)$ $i. e.$ $\phi \sim G^{-1}$. The importance of Brans - Dicke theory lies beyond its level of simplicity and it have strong theoretical supports from a variety of perspectives. For example  i) Brans - Dicke theory validates Mach's principle ii) it involves the simplest form of non - linear kinetic terms for scalar field which can be expected to serve as a very relevant candidate to play important role in the late time accelerated expansion of the Universe and some useful applications of Brans - Dicke theory are given in Refs. \cite{Green/1987,Appelquist/1987,Dvali/2000,Nicolis/2009,Deffayet/2011,Fujii/2003,Yadav/2020prd,Akarsu/2020,Avilez/2014}. Note that in Sen and Sen \cite{Sen/2001}, it has been argued that the late time acceleration is produced due to amalgamation of Brans - Dicke theory with small negative $\omega$ without inclusion of dark energy (DE) components in the matter/energy content of the Universe while in Bertolami et al. \cite{Bertolami/2000}, $\phi^{2}$ potential with large $\omega$ does not favor the positive energy condition for matter and scalar field both. In the recent past, Chimento \cite{Chimento/2010} has studied linear and nonlinear interactions in the dark components of the Universe with energy transfer.

In 2013, Tsallis and Cirto \cite{Tsallis/2013} have introduced the horizon entropy of a black hole as $S_{\delta} = \gamma A^{\delta}$ with $A$ is the area of Horizon; $\delta$ denotes the non-additivity parameter and $\gamma$ is unknown constant. Later on by assuming Hubble horizon as the IR cutoff in agreement of the thermodynamic considerations given in Ref. \cite{Jahromi/2018} and Tsallis entropy \cite{Tsallis/2013}, Tavayef et al. \cite{Tavayef/2018} have investigated a holographic dark energy model in the framework of General Relativity which is called Tsallis holographic dark energy (THDE) model. It is worthwhile to note that the Bekenstein entropy can be obtained by applying the Tsallis statistics to the system horizon \cite{Majhi/2017,Abe/2001,Biro/2011}. In Rashki and Jalalzadeh \cite{Rashki/2015}, authors have investigated that Tsallis entropy \cite{Tsallis/2013} of the system is power law function of its area by using the concept of quantum gravity. In Ref. \cite{Nunes/2016}, we see some useful cosmological features of Tsallis entropy. Some other recent works on THDE in the framework of Brans - Dicke theory of gravity are also available in the Refs. \cite{Ghaffari/2018,Aditya/2019}. However, in Xu et al. \cite{Xu/2009}, authors have investigated that the Hubble horizon as an IR cutoff in Brans - Dicke theory is not a suitable candidate to explain the late time accelerated expansion  of the Universe.

The article is organized as follows: Section \ref{sec:2} deals with the theoretical model and its basic mathematical formalism. In section \ref{sec:3}, the violation/validation of energy conservation law for THDE in Brans- Dicke cosmology are discussed. Finally, in section \ref{sec:4}, the article ends with discussion and final remarks.
\section{Theoretical model and \textbf{basic} equations}
\label{sec:2}
The modified Einstein-Hilbert action for Brans-Dicke theory of gravity is read as
\begin{equation}
\label{A-1}
S = \int d^{4}x\sqrt{g}\left(-\varphi R + \frac{\omega}{\varphi}g^{ij}\partial_{i}\varphi\;\partial_{j}\varphi+\mathcal{L}_{m}\right),
\end{equation}
where $\mathcal{L}_{m}$ and $\omega$ denote the matter Lagrangian density and dimensionless Brans-Dicke coupling parameter between scalar field and gravity respectively. The other symbols have their usual meaning.

On re-defining the scalar field $\varphi$ as
\begin{equation}
\label{A-2}
\varphi = \frac{\phi^{2}}{8\omega}.
\end{equation} 
Thus, the action (\ref{A-1}) in canonical form is recast as \cite{Setare/2017,Arik/2006,Arik/2008}
\begin{equation}
\label{A-3}
S = \int d^{4}x\sqrt{g}\left(-\frac{\phi^{2}}{8\omega} R + \frac{1}{2}g^{ij}\partial_{i}\phi\;\partial_{j}\phi+\mathcal{L}_{m}\right).
\end{equation}
The homogeneous and isotropic FRW metric is read as
\begin{equation}
\label{E-1}
ds^{2} = -dt^{2}+a(t)^{2}\left(\frac{dr^{2}}{1-kr^{2}}+r^{2}d\Omega^{2}\right),
\end{equation}
where $a(t)$ is the sacle factor which measures the rate of expansion of the Universe and $k =-1$, $0$ \& $1$ denotes the curvature parameter corresponding to open, flat and closed Universe respectively. We consider that the current Universe is filled with Dark Matter (DM) and DE. The pressure and energy density of DM are zero and $\rho_{M}$ while the pressure and energy density of DE are $p_{D}$ and $\rho_{D}$ respectively.

Thus, the field equations for the Universe filled with DM and DE are obtained as 
\begin{equation}
\label{E-2}
\frac{3}{4\omega}\phi^{2}\left(H^{2}+\frac{k}{a^{2}}\right)-\frac{\dot{\phi}^2}{2}+\frac{3}{2\omega}H\phi\dot{\phi} = \rho_{m}+\rho_{D},
\end{equation}
\[
-\frac{\phi^{2}}{4\omega}\left(2\frac{\ddot{a}}{a}+H^{2}+\frac{k}{a^{2}}\right)-\frac{1}{\omega}H\phi\dot{\phi}-\frac{1}{2\omega}\ddot{\phi}\phi \;-\;
\]
\begin{equation}
\label{E-3}
\frac{\dot{\phi}^{2}}{2}\left(1+\frac{1}{\omega}\right) = p_{D}.
\end{equation}
Here, an over dot denotes the derivative with respect to time and $H = \frac{\dot{a}}{a}$ is the Hubble parameter. In addition, the wave equation for scalar field is given by\\
\begin{equation}
\label{E-4}
\frac{\ddot{\phi}}{\phi}+3H\frac{\dot{\phi}}{\phi}-\frac{3}{2\omega}\left(\frac{\ddot{a}}{a}+H^{2}+\frac{k}{a^{2}}\right) = 0.
\end{equation}
Note that Eqs. (\ref{E-2}) - (\ref{E-4}) in this comment are same as Eqs. (2) - (4) in Ghaffari et al. \cite{Ghaffari/2018}.

It is worthwhile to note that Eqs. (\ref{E-2}) - (\ref{E-4}) are not closed because it is a system of three equation with four variables $H$, $\phi$, $\rho_{M}$ and $\rho_{D}$. Therefore one can not solve these equations in general. To get the explicit solution, we have to assume one physically viable condition which satisfies the above equation. For this sake, the authors of Ref. \cite{Ghaffari/2018} have assumed the scalar field $\phi$ as the power function of scale factor $a$ $i. e.$
\begin{equation}
\label{pl}
\phi = a^{n},
\end{equation} 
where $n$ is an arbitrary constant.

To construct the realistic model of the Universe, one has to obtain relation between $\phi$ and $a$, either by solving Eq. (\ref{E-4}) in general or assumed relation should follow Eq. (\ref{E-4}) for all or some selected values of free parameter. \\
Now, from Eq. (\ref{pl}), we obtain
\begin{equation}
\label{pl-1}
\frac{\dot{\phi}}{\phi} = nH,
\end{equation}
\begin{equation}
\label{pl-2}
\frac{\ddot{\phi}}{\phi} = n^{2}H^{2} + n\dot{H}.
\end{equation}
Using Eqs. (\ref{pl-1}) and (\ref{pl-2}), Eq. (\ref{E-4}) leads
\begin{equation}
\label{pl-3}
n\dot{H}+(n^{2}+3n)H^{2} = \frac{3}{2\omega}\dot{H}+\frac{3}{\omega}H^{2}+\frac{3}{2\omega}\frac{k}{a^{2}}.
\end{equation} 
To validate Eq. (\ref{pl-3}), we have the following
\begin{equation}
\label{pl-4}
n = \frac{3}{2\omega},\;\; n^{2}+3n = \frac{3}{\omega}\;\;\&\;\; k = 0.
\end{equation}
Now, we get
\begin{equation}
\label{pl-5}
n = -1,\;\; \& \;\; \omega = -\frac{3}{2}.
\end{equation}
Thus, we observe that in Brans-Dicke cosmology, $\phi \propto a^{n}$ is not general and its validation depends upon the some specific choice of free parameters. But in Ref. \cite{Ghaffari/2018}, authors have fixed $\omega = 1000$ and $n = 0.001$ to describe various features of the Universe. This hypothetical assumptions are not able to describes the whole aspects of Tsallis holographic DE in Brans-Dicke cosmology. So, in principle there is no compelling reason for the choice of $\phi \propto a^{n}$. However, we do not rule out the possibility to construct a toy model of the Universe filled with DM and DE with this assumption in Brans-Dicke cosmology. It is worthwhile to mention that $n < 0$ gives a flip from positive to negative value of deceleration parameter and the Universe shows a transition from deceleration to acceleration \cite{Banerjee/2007}. Further, we notice that the energy conservation law for THDE in Brans-Dicke gravity does not hold $i. e.$ the energy conservation equation (9) of targeted paper \cite{Ghaffari/2018} is not true for the values of $\rho_{D}$, $\dot{\rho_{D}}$ and $\omega_{D}$ given in Eqs. (7), (11) and (15) of Ghaffari et al. \cite{Ghaffari/2018} respectively. 
\section{Energy conservation law for THDE}\label{sec:3}     
In this section, firstly we discuss the energy conservation law for DM and DE in the framework of General Relativity (GR). The well known Einstein's field equation is read as
\begin{equation}
\label{g-1}
R_{ij}-\frac{1}{2}Rg_{ij}=-(T_{ij}^{m}+T_{ij}^{D}).
\end{equation}
Thus, the field equations for metric (\ref{E-1}) are given by
\begin{equation}
\label{g-2}
3\frac{\dot{a}^{2}}{a^{2}} + 3\frac{k}{a^{2}} = \rho_{m}+\rho_{D},
\end{equation}
\begin{equation}
\label{g-3}
2\frac{\ddot{a}}{a}+3\frac{\dot{a}^{2}}{a^{2}} +\frac{k}{a^{2}} = -p_{D}.
\end{equation}
Differentiating Eq. (\ref{g-2}) with respect to t, one can express the resulting answer as a linear combination of Eqs. (\ref{g-2}) and (\ref{g-3}) as following
\begin{equation}
\label{g-4}
\dot{\rho_{m}}+3\rho_{m}H +\dot{\rho_{D}}+ 3\left(\rho_{D}+p_{D}\right)H = 0.
\end{equation}
Eq. (\ref{g-4}) is a direct consequence of the energy conservation law implicit in the Einstein equations \cite{Narlikar/2002}. It is worthwhile to note that one can not obtain Eq. (\ref{g-4}) from Eqs. (\ref{E-2})-(\ref{E-4}) by solving it in similar way. So, the non conservation of energy in Brans - Dicke cosmology raises a serious question about the significance of the energy momentum tensor in this theory. The satisfactory answer of this question had been given in Ref. \cite{Brans/1961}. In Brans - Dicke \cite{Brans/1961}, the author have assumed that the matter Lagrangian density $\mathcal{L}_{m}$ is a function of matter variables and $g_{ij}$ only and not a function of $\phi$ (see Ref. \cite{Brans/1961} for detail). To make it clear, here, we have re-written the field equations for the Universe filled with DM and DE in Brans - Dicke cosmology in general form as following \\
\begin{equation}
\label{ef-nH1}
2\frac{\ddot{a}}{a}+\frac{\dot{a}^{2}}{a^{2}} +\frac{k}{a^{2}}  = -\frac{p_{D}}{\phi} -\frac{\omega}{2}\frac{\dot{\phi}^{2}}{\phi^{2}}-2\frac{\dot{\phi}}{\phi}\frac{\dot{a}}{a} - \frac{\ddot{\phi}}{\phi} = -p^{(eff)},
\end{equation}
\begin{equation}
\label{ef-nH2}
3\frac{\dot{a}^{2}}{a^{2}} + 3\frac{k}{a^{2}} =  \frac{\rho_{D}}{\phi}+\frac{\rho_{m}}{\phi} + \frac{\omega}{2}\frac{\dot{\phi}^{2}}{\phi^{2}}-3\frac{\dot{\phi}}{\phi}\frac{\dot{a}}{a} = \rho^{(eff)},
\end{equation}
where $p^{(eff)} = p_{m}+p_{D}+p_{\phi}$ and $\rho^{(eff)} = \rho_{m}+\rho_{D}+\rho_{\phi}$. Note that $p_{\phi}$ and $\rho_{\phi}$ stand for pressures and energy densities due to Brans - Dicke scalar field \cite{Amirhashchi/2019,Yadav/2020}.\\

Differentiating Eq. (\ref{ef-nH2}) with respect to time, one can express the resulting answer as a linear combination of Eqs. (\ref{ef-nH1}) and (\ref{ef-nH2}). Therefore
\begin{equation}
\label{ef-nH3}
\dot{\rho}^{(eff)}+3(\rho^{(eff)}+p^{(eff)})H = 0.
\end{equation} 
The above equation is the direct consequence of the energy conservation law in Brans-Dicke gravity. It is worthwhile to note that Eq. (\ref{ef-nH3}) is true only for scalar field independent Lagrangian density. But in Ref. \cite{Ghaffari/2018}, the energy density of THDE with Hubble radius as IR cutoff is defined as 
\begin{equation}
\label{T-1}
\rho_{D} = B\phi^{2\delta}H^{4-2\delta}.
\end{equation}   
Eq. (\ref{T-1}) is same as Eq. (7) of targeted paper \cite{Ghaffari/2018}. In Ghaffari et al. \cite{Ghaffari/2018}, B is not defined and $\delta$ is taken as arbitrary constant. However, we understand that $B$ is unknown parameter. From Eq. (\ref{T-1}), we observe that $\rho_{D}$ is function of $\phi$ which may lead \textbf{to} the violation of energy conservation law for THDE in Brans -Dicke cosmology.
\section{Discussion and final remarks}\label{sec:4} 
In this comment we investigate that the power law variation of scalar field with scale factor for the Universe filled with DM and DE is restricted for some particular values of free parameters. We have obtained the values of free parameters $n$ and $\omega$ by bounding $\phi \propto a^{n}$ with wave equation of scalar field. This values of $n$ and $\omega$ do not match with its values assumed in Ghaffari et al. \cite{Ghaffari/2018}. It is interesting to note that a diiferent approach for bounding the values of $n$ and $\omega$ are given in Ref. \cite{Banerjee/2007}. Banerjee and Pavon \cite{Banerjee/2007} have considered holographic DE and $\phi \propto a^{n}$ in Brans - Dicke gravity and investigated a case of particular interest that is for small $\mid n \mid$, $\omega$ is high so that the
product $n^{2}\omega$ results of order unity while in Ghaffari et al. \cite{Ghaffari/2018}, these values are conveniently assumed. Also, we observe that $n = 0$ and consequently $\omega = \infty$ are solutions of Eq. (\ref{pl-4}). These solutions are noticeable because $n = 0$ leads to $\phi = 1$ and hence Eqs. (\ref{ef-nH1}) and (\ref{ef-nH2}) are reducing to Eqs. (\ref{g-2}) and (\ref{g-3}) respectively. This is of course the case of general relativity. The standard belief exposed in the textbooks \cite{Narlikar/2002,Weinberg/1972} is that, as $\omega \rightarrow \infty$, the Brans - Dicke theory tends to general relativity. Another problem in Ghaffari et al. \cite{Ghaffari/2018} is that this model does not hold the energy conservation law for THDE as we have discussed in section 3. 

It is worthwhile to note that neither we avoid the variation of scalar field as power function of scale factor nor decline the possible existence of THDE in Brans - Dicke gravity. Regarding $\phi \propto a^{n}$, the bounds on free parameter $n$ and $\omega$ should be clearly identified. Secondly the violation of energy conservation law for THDE in Brans - Dicke cosmology raises a serious question on the method and approach given in Ref. \cite{Ghaffari/2018}. As a final comment, we note that in spite of good possibility of THDE model in Brans - Dicke gravity to render a theoretical foundation for cosmology, the experimental point is yet to be weighed and still the theory needs a fair trial. \\
\section*{Acknowledgements}
The author is grateful to the reviewer for useful comment. During preparation of this manuscript, we have approached S. Ghaffari with some queries and we appreciate S. Ghaffari for his inputs.

\end{document}